# Concept-Oriented Programming: References, Classes and Inheritance Revisited

Alexandr Savinov

Database Technology Group, Technische Universität Dresden, Germany
http://conceptoriented.org

**Abstract**

The main goal of concept-oriented programming (COP) is describing how objects are represented and accessed. It makes references (object locations) first-class elements of the program responsible for many important functions which are difficult to model via objects. COP rethinks and generalizes such primary notions of object-orientation as class and inheritance by introducing a novel construct, concept, and a new relation, inclusion. An advantage is that using only a few basic notions we are able to describe many general patterns of thoughts currently belonging to different programming paradigms: modeling object hierarchies (prototype-based programming), precedence of parent methods over child methods (inner methods in Beta), modularizing cross-cutting concerns (aspect-oriented programming), value-orientation (functional programming). Since COP remains backward compatible with object-oriented programming, it can be viewed as a perspective direction for developing a simple and natural unified programming model.

## 1. Introduction

Object orientation is one of the most influential and successful paradigms in computer science. Much of its power comes from the ability to express a wide variety of real-world situations using only a few relatively simple concepts the main of which is that of *object*. Objects have always been in the center of this methodology (hence its name) according to which it is object's functionality that accounts for most of the program complexity. In other words, a program is a number of objects possessing some behavior and responding to messages, and to describe a program means to describe objects it consists of.

Although object-oriented programming (OOP) has been proven to be extremely successful, it has one general drawback: it does not provide a means for describing how objects are *represented* and how they are *accessed*. Strictly speaking, it is not a drawback but rather a consequence of its main assumption that everything should be described by objects. As a result, the mechanism of object representation and access is supposed to be provided by the translator. Any object is guaranteed to get some kind of primitive reference and a built-in access procedure without a possibility to change them. Thus there is a strong asymmetry between the role of objects and references in OOP: objects are intended to implement domain-specific structure and behavior while references have a primitive form and are not modeled by the programmer. Programming means describing objects but not references.

One reason for this asymmetry is that there is a very old and very strong belief that it is entity that should be in the focus of modeling (including programming) while identities simply serve entities. And even though object identity has been considered "a pillar of object orientation" [Ken91] their support has always been much weaker in comparison to that of entities. Another reason is that hiding the structure of references and the design of access procedures is a very successful pattern which, particularly, accounts for the success of OOP. Indeed, we are interested only in what this object does rather than how it is being accessed, and we want to abstract from the representation and access details when manipulating objects. However, this abstraction from reference mechanics is achieved by sacrificing the possibility to model these functions at all by completely removing references and object access procedures from the scope of programming and delegating these functions to the translator. In OOP, we are not able to model how objects exist, where they exist, and how they are accessed.

Concept-oriented programming (COP) (first described in [Sav05a]) is a novel approach to programming the main general goal of which is to answer these questions by *legalizing references* and making them first-class elements of programming languages. In this sense, COP can be characterized as reference-oriented programming or programming focusing on what happens during access. It is assumed that references account for a great deal of the program complexity and their functions are at least as important as those of objects. For example, in the following typical object-oriented program we create an object representing a bank account and then call its method:

```
Account acc = new Account();
acc.setInitialBalance(100.00);
```

In OOP, it is not possible to specify our own custom format of references for the bank account object although it would be natural to use a domain-specific format like account number. And we are not able to provide our own access procedure even though it is natural that any access to this object has to be somehow controlled. All variables store only platform-specific (primitive) references and the translator creates the illusion of *instantaneous* object access without any possibility to specify domain-specific intermediate actions.

COP changes this view by making references *active* elements of the program with arbitrary structure and behavior. For example, bank accounts can be represented directly by account numbers and have functions responsible for access to the account object. To describe both references and objects, COP introduces a novel construct, called *concept* (hence the name of this approach). The main goal of concepts is to retain main functions of conventional classes by providing a possibility to model how objects are represented and accessed. In previous versions of COP [Sav08a, Sav09a, Sav08b], concept was defined as a *couple* of two classes: one object class and one reference class. For example, a bank account concept is defined as follows:

```
concept Account {
  reference { // Reference class
    char[10] accNo;
  }
  object { // Object class
    double balance;
  }
}
```

Since references and objects are symmetric, programming can be viewed as consisting of two orthogonal branches. On one hand, we can model references and their behavior. On the other hand, we can model objects.

An important point is that reference and object classes are defined only together, that is, a concept has one name for both its constituents. When concepts are used, they are undistinguishable from classes. For example, a variable of the `Account` concept is declared as usual:

```
Account acc;
```

In contrast to OOP, it will always store a reference in the format specified by in the reference class of the concept while the object is accessed indirectly using this concept access methods. If reference class is empty then it will be provided by the translator so we get conventional classes. If object class is empty then such concepts describe values. In the general case, the programmer can freely vary what part of an element is passed by-value and by-reference.

Classical inheritance cannot be easily adopted for concepts, particularly, because concept instances exist in a hierarchy (like objects in prototype-based programming). Therefore a new relation was introduced, called *inclusion*. Its main purpose consists in modeling hierarchical address spaces by describing references consisting of several segments. As a result, objects in COP exist in a hierarchal space where each of them has a unique address with custom structure (like postal addresses). Defining program elements as consisting of two parts and existing in a hierarchical address space leads to rethinking and generalizing such fundamental notions as object identification, inheritance and polymorphism.

First version of concept-oriented programming is described in [Sav05a] and is denoted as COP-I. The next version, described in [Sav08a, Sav09a, Sav08b] and denoted as COP-II, changes the interpretation of concepts and adds several new mechanisms. This paper is a full version of the paper [Sav12a] describing a new major revision of concept-oriented programming, denoted as COP-III. The main goal of the third revision is to describe this programming model by using fewer general notions and more natural interpretations by simultaneously covering more programming patterns existing in other approaches.

The first major change in COP-III is that concepts are defined differently: instead of using two constituents – object class and reference class – we use only one constituent which models references. Thus objects are effectively excluded from the model as a structural element. If COP-I and COP-II treat references and objects as two symmetric constituents then COP-III makes references more important than objects. Yet, objects are still fully supported in the model. Instead of modeling them explicitly via object classes, we propose a new general treatment: *objects are functions of their references*. In COP-III, this definition is made more specific: objects are defined via outgoing methods of concepts. Thus programming in COP-III is reduced to describing locations (references) and their functions (objects). For modeling duality, COP-III uses two kinds of methods – incoming methods and outgoing methods – instead of using two kinds of classes – reference class and object class.

Another important change is the use of two keywords for navigating through the hierarchy, super and sub (as opposed to using only super in OOP), and the existence of two opposite overriding strategies. This allows us to treat any element as a domain, scope or context consisting of many internal child elements. Incoming methods of concepts intercept requests from outside and outgoing methods intercept requests from inside.

COP-III also removes the reference resolution mechanism and continuation method from the programming model. Instead, access indirection relies on the ability of elements to intercept incoming and outgoing methods.

The paper has the following layout. Section 2 provides general background and motivation for COP by describing its design goals and basic principles. Section 3 defines the notion of concept and how concepts are used for modeling objects and references. Section 4 is devoted to describing inclusion relation and how it is used for modeling address spaces and object hierarchies. Section 5 describes how inheritance, polymorphism and cross-cutting concerns are implemented in COP-III. Section 6 provides some discussion with an overview of related work and Section 7 makes concluding remarks.

## 2. Background and Motivation

### 2.1 General design goals

*Modeling values and objects.* By *values* we mean data passed and stored directly by copying their contents while *objects* are data passed and stored indirectly (by-reference) using a value serving as its representative. Conventional OOP focuses on objects while functional programming (FP) has a strong support of values. Values describe what is transferred and exists transiently (data packets, method parameters, messages etc.) while objects describe shared data that exists persistently. A problem of conventional classes (as well as many other similar constructs) is that they do not allow us to distinguish between objects and values. One existing solution is to use two different constructs for describing values and objects like `struct` for values and `class` for objects. Another solution is to provide this specification for each variable, parameter or field using some keywords like `byref` and `byval`. However, we believe that both values and objects are equally important for programming and the goal is to make them symmetric elements with equal rights being modeled using *one* construct.

*Modeling references.* *References* are values which manifest the fact of object's existence. Their main functions consist in distinguishing objects, representing and providing access to them. Most programming languages provide only primitive references which cannot be extended by modifying their structure and behavior, nor is it possible to define new custom references. In this case all objects exist in one flat space under control of one manager with very limited possibilities to influence its behavior. If we need some specific object manager with domain-specific structure of references and functions then it has to be developed manually without any support from the language just because these notions (reference, object access, object life-cycle etc.) are not well supported. However, we believe that references are responsible for a great deal of domain-specific functions and therefore they should be at least as important as objects. In other words, a program can be written in terms of objects but it also can be written in terms of references and having weak support for the latter results in using complex workaround techniques and patterns like smart pointers in C++ or proxies. Our goal in this context is to legalize references by directly supporting them and giving them at least the same status as that for objects.

*Modeling hierarchical addresses.* Custom references provide a mechanism for defining arbitrary domain-specific address spaces where objects exist. However, such an address space is not structured so that objects of one class still exist in a flat space. Managing a number of flat spaces is not very convenient and the goal here is to support hierarchical address spaces where each object is represented by a complex reference consisting of several segments. Each next segment of such a reference is a relative address with respect to the previous segment which is similar to postal addresses.

*Modeling object hierarchy.* Most class-based approaches are characterized by asymmetry between classes and their instances: classes exist in a hierarchy while their instances exist in flat space. This means that parent classes are shared among their child classes while each child object has its own parent object which is not shared. Thus class-based approaches allow us to model class hierarchies but not object hierarchies. We can re-use behavior of parent objects but not data. Prototype-based programming provides means for modeling object hierarchies but without classes. The goal of COP is to eliminate this asymmetry by supporting object hierarchies which are modeled by class hierarchies.

*Access indirection and object protection.* In OOP, object access via method calls is treated as an instantaneous action which means that it is not possible to intervene into this process. Hiding the implementation details of the mechanism of method calls is of course a good feature. What is bad is that we cannot adapt this procedure to the purposes of this concrete program by implicitly executing some intermediate domain-specific code for each method call. The mechanism of access indirection and interception can be implemented manually using various patterns like proxy. Another approach is using inner methods [Gol04]. However, these approaches do not support object hierarchies. Our goal is to make access indirection and interception inherent features of objects so that any object can easily control all incoming and outgoing accesses. The logic of intermediate processing and interception should be part of the object functionality so that an object is treated as a space border with the logic of intermediate processing. In this case an object in a hierarchy can be represented as a *set* of its child objects where it plays a role of space border. This border should be controlled by intercepting all access requests not only to this but also to all child objects. We regard border control as a general property of any system defined as a set of its internal elements. If we want objects to play a role of a system then they should be able to intercept all incoming method calls with the purpose to protect itself and child objects.

*Injection and enforcing behavior.* In OOP, objects can re-use the behavior of other objects but there is no way to force other objects to behave in certain way. In other words, it is not possible to inject functionality into other objects by changing their behavior without their explicit desire to do so. This feature would be especially desirable in the case of object hierarchies where one parent object could modify behavior of all its child objects. For example, if a `Panel` object is a set of `Button` objects (a panel consists of several buttons) then the panel might want to enforce some common behavior like drawing its border or background as a *necessary* function that is executed im-plicitly and cannot be removed (so that the child buttons are even not aware of this intervention). Another example is where a parent object needs to check access permissions before this method call is passed further to the child object. Obviously, this behavior must be enforced because voluntary security checks (by calling parent methods from children) hardly make sense. This feature is very similar to what aspect-oriented programming (AOP) is intended for [Kic97]. However, aspects in AOP are orthogonal to classes while we would like to make injection an inherent feature of classes so that injection is aligned with the class hierarchy. A similar approach is based on using inner methods [Gol04] which however does not support object hierarchies.

## 2.2 General principles of COP

*Objects in space.* In COP, it is important to think of objects as being located in a structured space (Fig. 1). What is even more important is that spaces are represented by normal objects, that is, an object *is* a space (of its internal objects) and an object is *in* a space (of its external object). Essentially, this means that we have only objects connected via set membership relation. One object can be a member of some space and include other objects as its members. It is assumed also that the space of objects has a nested structure where one object is included in one parent object. It can be represented as a tree of objects which is analogous to prototype-based programming. The difference is in how a tree of objects is interpreted and implemented. COP uses a tree of classes for modeling a tree of objects. Other differences are in how objects are represented which is described below.

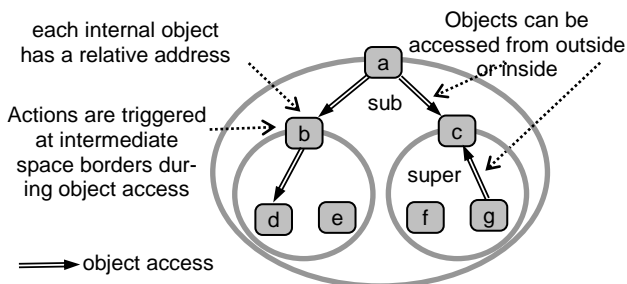

**Fig. 1.** Concept-oriented program is a nested space

*Objects address.* Each object has a unique address the structure of which corresponds to the structure of space where it exists. An address is composed of several segments where each segment is a local address with respect to the parent space. Essentially, describing the address structure is equivalent to describing the structure of the space. For example, four leaf objects in Fig. 1 are represented by addresses <a,b,d>, <a,b,e>, <a,c,f> and <a,c,g>. Importantly, objects may get arbitrary domain-specific addresses and their structure is part of the object type description. Moreover, addresses possess some behavior which describes how objects are accessed. Having domain-specific addresses with behavior is a new feature. The change of paradigm is that the overall structure and behavior of a program is now distributed between two constituents: objects and addresses (references). In contract, the conventional approach assumes that they belong to only objects.

*Object access.* Interactions in such a system cannot propagate instantly because objects have different locations in the hierarchy (in space). Note that objects have domain-specific addresses and can be located anywhere in the world. For example, an object might well have a reference coinciding with a postal address. In this case accessing an object can be a rather complicated procedure consisting of many steps where one step corresponds to one border intersection.

*Interception at space borders.* Objects in the system act as space borders by automatically intercepting all incoming and outgoing access requests. These intermediate functions are triggered automatically and it is not possible to access an object without such an interception if the source and the target objects are separated by another object. For example, the object <a,b,d> can be accessed from the object <a> only by crossing the intermediate border represented by the object <a,b> which will intercept such a request and perform arbitrary actions. The change of paradigm is that intermediate functions executed automatically during object access account for a great deal or even of the system complexity.

## 3. Concepts instead of Classes

### 3.1 Concepts and values

Value types in COP are modeled by means of *concepts* where concept fields describe the structure of values and concept methods describe their behavior. In this sense, concepts are equivalent to conventional classes (particularly, to classes in C++) except that their instances are always values and there is no direct way to produce an object from a concept, for example, by getting its address. Concept instances are passed by-copy only and do not have any permanent location, address, pointer, reference or any other representative. For example, the following concept describes a bank account:

```
concept Account {
  char[10] accNo;
  Person owner;
}
```

The first field will contain 10 characters while the second field will contain a value with the structure defined by the `Person` concept.

### 3.2 Dual methods

What makes concepts different from classes is the presence of two kinds of methods: *incoming* methods and *outgoing* methods. Such a pair of incoming and outgoing methods with the same signature is referred to as *dual* *methods*. Incoming methods will be marked by the modifier 'in' while outgoing methods will be marked by the modifier 'out'. For example, if we would like to have a method for getting the current account balance then this functionality can be specified in the incoming and/or outgoing methods:

```
concept Account
  char[10] accNo;
  in double getBalance() {...};
  out double getBalance() {...};
}
```

It is not necessary to define both versions: if one of them is absent then it is supposed to have a default implementation.

Although there are two definitions for each method, they are still used as one method. In other words, methods are called as usual using only their name without any indication if it is an incoming or outgoing version. The main purpose of dual methods is performing different actions for different scopes and directions of access. If concepts describe borders then there are different implementations for incoming and outgoing requests. In other words, a request originating from inside is processed by an outgoing method and a request originating from outside is processed by an incoming method.

### 3.3 Objects and references

Concepts can be easily used to model references because references *are* values. More specifically, references are values providing access to other values which are thought of as stored in object fields. Thus object fields can be defined as functions of references. Formally, if values are represented by tuples consisting of other values then objects are represented by tuples consisting of functions. Value tuples are written in angle brackets and function tuples (objects) are written in parentheses. Functions are defined in terms of the corresponding reference (value tuple) and therefore an element is COP is defined as a *couple* consisting of one value tuple (reference) and one function tuple (object):

$$[\text{element}]\ s = \langle v_1,\ldots,v_n\rangle(f_1(v_1,\ldots,v_n),\ldots,f_m(v_1,\ldots,v_n))\quad(1)$$

Importantly, only the reference part of an element is really transferred while the object part is what the functions return.

It is rather general definition which means that values are interpreted as locations and then some functions of these locations return values interpreted as being stored in object fields. We do not say anything about object allocation or storage for object fields. It is assumed that given a reference (a value) one can always get some additional values using the associated functions as if these values were *stored* in the object. However, we do not know where really these values are stored and if they are stored at all (they could be computed or requested from some service). Using references and outgoing methods as object fields creates the illusion that some values are really stored at this location. However, the value can be stored anywhere while the reference is needed to read or write it. References are thought of as virtual addresses rather than direct pointers to object fields. For example, URLs are virtual addresses but we still think of the web pages as being stored at these locations even though their content could be generated or stored on many computers.

In the previous versions of COP, it was assumed that an element is a couple. However, it was a couple of two explicitly defined tuples, that is, both the reference part and the object part had their own structure. In the new version, only the reference part has an explicit structure while the object part is defined via functions. Thus object as an independent construct is removed from the model. Instead, objects are being modeled using references and their functions.

The next step in object modeling mechanism is to assume that *object fields are implemented via outgoing methods of concepts* which return the same result for the same reference. Syntactically, we will define such methods as setters and/or getters. For example, bank accounts are uniquely identified by their numbers which are used as a reference. In addition, any bank account is supposed to have a balance which however should be stored in an object field. Such a field is defined using an outgoing method which returns the balance depending on the account number.

```
concept Account {
  char[10] accNo;
  out double balance {
    get { return func(accNo); }
  }
}
```

Here we effectively defined a new object field, called `balance`, which can be used as usual:

```
Account acc = getAccount("Smith");
double currentBalance = acc.balance;
acc.balance = acc.balance + 100.0;
```

In the case of no hierarchy, outgoing fields can be implemented by the compiler using the functionality provided by primitive references: For that purpose, we need to simply mark a field as an outgoing member and also extend a primitive reference (see next section for more information about inclusion relation):

```
concept Account extends MemoryAddress {
  char[10] accNo;
  out double balance;
}
```

The compiler will normally allocate some memory for the object and then use the reference for access to its fields. This declaration is analogous to the treatment of concepts in previous versions of COP where reference members and object members were explicitly separated:

```
concept Account // COP-I and COP-II
  reference {
    char[10] accNo; // By-value
  }
  object {
    double balance; // By-reference
  }
```

# 4. Inclusion Instead of Inheritance

## 4.1 Hierarchical address spaces

A concept can be *included* in another concept and this relation is denoted by the keyword 'in'. If concept B is included in concept A then instances of B will extend instances of A. In this sense, concept inclusion describes conventional value extension.

What is new in inclusion is that it describes hierarchical address spaces and space inclusion. Here we use an important conceptual assumption that if an address is a value then an *extended more specific value is a relative address*. In this case parent concepts describe spaces for their child concepts while concept fields define the structure of local addresses in the parent space. A child instance (extension) is said to exist in the domain, context or scope of its parent instance (base).

For example, bank accounts are always identified with respect to their bank. Such a hierarchical space is modeled by two concepts and inclusion relation:

```
concept Bank {
  char[12] bankCode;
}

concept Account in Bank {
  char[10] accNo;
}
```

A reference to an account object will consist of two segments: a parent bank reference and a child account reference.

Concepts can be included in a primitive reference provided by the platform or library like global heap, local heap, remote reference or persistent storage. In this case, it can rely on its functions which are normally used to implement object fields. For example, if we are going to allocate our objects in memory then we use the standard memory manager:

```
concept Bank in MemoryHandle {
  char[12] bankCode;
}
```

Now instances of the Bank and Account concepts will extend memory handles provided by the platform. By default (but not always), each new bank and account objects will get a separate memory handle. In particular, each variable of the Account concept

```
Account acc; // 3 segments
<mem:bank:acc>
```

will consist of three segments: memory handle, bank code and account number. (The compiler will allocate memory handles and the size of memory necessary to store all object fields.)

## 4.2 Navigating a concept inclusion hierarchy

A concept breaks the whole space into two domains: internal and external. Internal domain consists of all its child concepts while external domain consists of all other concepts. If we think of a concept as a border (Fig. 2) then it can be crossed in two directions: from outside in the direction of internal domain and from inside in the direction of external domain. Note that method calls do not provide any indication whether they are incoming or outgoing calls. The rule is that if an element is accessed from inside then its outgoing methods are used, and if it is accessed from inside then its incoming methods are used. Concepts provide two implementations for each method: one for external use and one for internal use. Once two versions of a method have been defined, we can forget about their differences and use concept methods precisely as methods of conventional classes. This can be also thought of as a visibility rule where outgoing methods are visible from inside while incoming methods are visible from outside. It is analogous to the passport control system at airports where arriving and departing passengers pass through different procedures.

COP uses super and sub keywords to access the parent and child elements, respectively. Applying a method to the sub keyword will produce an incoming method call because we are trying to enter the domain (Fig. 2). Applying a method to the super keyword will call an outgoing method of the parent concept because it is a call from inside. Thus super method calls are always outgoing methods and sub method calls are always incoming methods. For example, if a method of the Bank concept is called from any method of the Account concept then an outgoing version of this method will be executed:

```
concept Account in Bank
  out double getInterest() {
    double bankRate = super.getInterest();
    return bankRate + accRate;
  }
}
```

Here super.getInterest() is an outgoing method of the Bank concept which returns the current interest rate at this bank. An incoming version of this method might produce different interest rate for external calls (or might not be defined at all). The getInterest method of the Account concept will be accessible from its child concepts only because it is marked as an outgoing method.

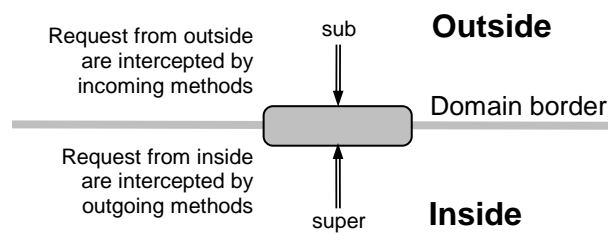

**Fig. 2.** Incoming and outgoing methods

## 4.3 Object hierarchies

One of the distinguishing features of COP is the support of object hierarchies where one object may have many child objects (with different relative references). In the previous versions of COP, each local reference was resolved into a *primitive* reference which was used to access object fields. Therefore, each child could resolve its reference to different primitive references where object fields are stored. In this current version we do not use the reference resolution function. Instead, all outgoing methods produce their result depending on this value and the parent values. In the case of the same parent, outgoing methods of different children will produce different results which are interpreted as different object field values. In this sense, each outgoing method acts similar to the reference resolution method used in the previous versions.

For example, assume that one bank object has many account objects the state of which is stored persistently in some database. Account balance could be then defined as follows:

```
concept Account in Bank {
  char[10] accNo;
  out double balance {
    get {
      Connection db = super.getConnect();
      return db.load("balance", accNo);
    }
    set {
      Connection db = super.getConnect();
      db.save("balance", accNo, value);
    }
  }
}
```

Here each `Account` object is identified by its number and then its `balance` object field is defined as an outgoing method (via one setter and one getter). The balance depends on the current bank (because it is a parent object which provides connection to the database). It also depends on the current account number (reference) because it is used as a primary key when getting values from the database.

Importantly, these are only implementation details but logically all objects exist in a hierarchy where each bank has many accounts. We can read balances and update balances using account references (consisting of several segments). And these operations will be logically correct because their result depends only on references.

# 5. Uses of the Inclusion Hierarchy

## 5.1 Inheritance

Inheritance is a language mechanism for defining new objects by *reusing* already existing object definitions. The most wide spread treatment of inheritance is that members of a new class are added to or extend those already defined in the base class being reused. Although this model is considered a particular case of inclusion, it can be easily implemented. The classical treatment of inheritance is directly supported by outgoing concept methods. This means that child outgoing methods are implemented using parent outgoing methods which are called via the `super` keyword.

Inheriting concept fields also works precisely as in the classical case: child fields are simply added to the parent concept fields. In this way we can extend values by adding more fields to them. For example, if concept Point has two fields x and y then we can define a new concept Point3D which has an additional field z:

```
concept Point { int x; int y; }
concept Point3D in Point { int z; }
```

Extending objects is not so simple because parent objects are shared among their children and therefore child fields cannot be simply concatenated with the parent fields. The classical model for object extension can be obtained if the child concept has no fields. Since the reference is empty, only one child can exist within one parent (just because they cannot be distinguished). In this case, we can think of child object fields as simply extending the parent fields. For example, if we need to define a bank account with some additional property then it can be done as follows:

```
concept BonusAccount in Account {
  out double bonus; // Object field
}
```

It is equivalent to conventional class and class inheritance (Fig. 3). Any instance of this class will get its own parent segment with an additional `bonus` field defined in this concept.

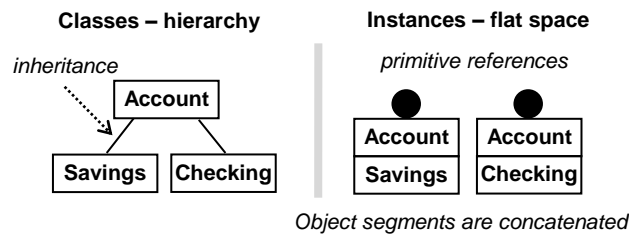

**Fig. 3.** Composition of objects in the case of OOP inheritance

## 5.2 Polymorphism and reverse overriding

Polymorphism allows an object of a more specific type to be manipulated generically as if it were of a base type. For example, if we declare a variable as having the type `Account` then polymorphism allows us to apply to it the method `getBalance` even though it stores a reference to a more specific type like savings account.

There exist different approaches to implementing polymorphic behavior but the currently dominating strategy consists in completely overriding parent methods by child methods. In other words, if we define a child method then it will have precedence over the parent method. If the child still needs some functionality provided by the parent method then it can access it using a super call. If a method is applied to a reference then the most specific implemen-tation is called first and then it can call more general versions of this method provided by its parents.

For example, if the `Button` class has to provide a more specific implementation of the `draw` method (than its parent `Panel` class) then it is usually implemented as follows:

```
class Panel {
  void draw() {
    fillBackground();
  }
}

class Button extends Panel {
  void draw() {
    super.fillBackground ();
    drawButtonText("MyButton");
  }
}
```

COP proposes to use a *reverse overriding strategy* for implementing polymorphism where parent methods have precedence over and then can call child methods. This strategy uses incoming methods for overriding. Thus incoming methods of parent concepts override incoming methods of child concepts. In our example, panel background is filled by the parent class and then the child method is called in order to add more specific behavior:

```
concept Panel {
  in void draw() {
    fillBackground();
    sub.draw();
  }
}

concept Button in Panel {
  in void draw() {
    drawButtonText("MyButton");
  }
}
```

Both strategies describe behavior incrementally by executing some operations and then sending a request for further processing either to the parent or child object so the difference between them is only in the direction of delegation (Fig. 4). The classical overriding strategy relies on parent methods when composing the necessary behavior. The reverse strategy relies on child methods to add more specific behavior from the extension. Note that these two strategies can be combined only if we have the mechanism of dual methods which effectively isolates two directions for method call propagation. Without dual methods it is easy to get an infinite cycle when parent and child versions call each other.

The main difference of the reverse overriding is that it *enforces* the behavior of parent classes so that parent methods are guaranteed to be executed. In other words, the parent object has always the final decision on what will be executed. In contrast, the classical approach does not guarantee that parent methods will be executed. For example, assume that the `Bank` concept wants to authorize all incoming requests before doing any specific actions by internal objects (extensions). In OOP it is impossible because the most specific implementation will be called first. The only way to overcome this difficulty is to stipulate this rule in documentation as a kind of discipline for developers of extensions. For example, this rule might say that when overriding base methods the developer should authorize the request by calling another base method. Obviously, this approach has numerous potential problems because it does not allow us to enforce the necessary logic. COP provides a principled solution where the developers can easily define any behavior which will be executed for each extension and cannot be overridden.

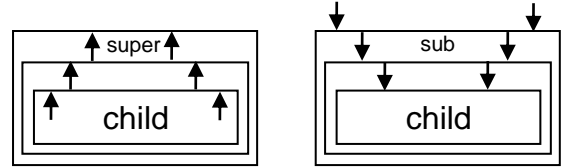

Child (specific) methods have precedence over parent methods and then use parent via super

Parent (general) methods have precedence over child methods and then use children via sub

**Fig. 4.** Two overriding strategies.

### 5.3 Interception and injection

Programs frequently have functions which are difficult to modularize because they are scattered throughout the whole source code. Such program logic that spans the whole program is referred to as a cross-cutting concern and it is known to produce numerous problems in software development. Several general solutions and specific mechanisms for this problem have been proposed like aspect-oriented programming (AOP) [Kic97], feature-oriented programming [Bat03], delegation-based mechanisms [Sch08] and context-oriented programming [Con05, Hir08].

COP proposes a novel solution for this problem which is based on the ability of parent methods to intercept any access to child methods. Thus cross-cutting concerns are modularized in parent incoming methods and this functionality is injected in child methods. Effectively, this mechanism allows using parent incoming methods as wrappers for child methods so that some functions are guaranteed to be executed for each access while target objects are unaware of this intervention. In terms of spaces, cross-cutting concerns are thought of as functions associated with borders. These functions are triggered automatically for each incoming request so that one and the same code is executed for many internal objects.

For example, if we would like to log any access from outside to account balances then this cross-cutting concern is implemented in the `getBalance` incoming method:

```
concept Bank {
  in double getBalance() {
    logger.Debug("Balance accessed.");
    return sub.getBalance();
  }
}
```

Note that behavior can be injected by several incoming methods as the method call propagates down along the inclusion hierarchy to the target.

Interestingly, the notion of cross-cutting concern can be also applied to outgoing methods which means that one and the same logic is executed for all outgoing requests. For example, if banks have some reserves and they want to log all accesses to this property from inside then it is implemented as an outgoing method:

```
concept Bank {
  protected out double reserves;
  out double getReserves() {
    logger.Debug("Reserves accessed.");
    return this.reserves;
  }
}
```

Now any access to the bank reserves from any child object (like `Account` methods) will be logged. Obviously, this pattern is easily implemented in OOP. We mention it in order to emphasize that cross-cutting behavior has dual nature which is implemented via incoming and outgoing methods.

## 6. Discussion and Related Work

### 6.1 Modeling values and references

Previous versions of COP, values and objects had the same rights. COP-III makes values the primary element of the model and therefore it can be called a value-oriented approach. This feature makes it closer to functional programming (FP) because, strictly speaking, COP manipulates only values. What is really new in our approach to values is that they can play a role of references. Values in COP are not simply some isolated data. Their main purpose is to *represent* other objects and therefore a value is treated as a *location* which has been the central notion in COP from the very beginning. As a consequence, the operation of extension which exists in many other approaches has a set-based interpretation: to extend a value means to build an internal location with respect to the base value treated as a set or domain. Thus all values (and the represented objects) exist within some domain, that is, we can always answer the question *where* one or another value exists.

### 6.2 COP vs. aspect-oriented programming

Aspect-oriented programming (AOP) [Kic97] is the most wide spread approach to modularizing cross-cutting concerns. AOP introduces an additional programming construct, called aspect, which has two major purposes: (i) it defines cross-cutting concerns (data and behavior, called advices), and (ii) it specifies points in the program where it has to be injected, called join-points. An important feature of AOP is that aspects are orthogonal to the class hierarchy so that behavior defined in aspects is then injected into points defined in the class hierarchy (Fig. 5). In this sense, aspects and classes play different roles; they are not completely unified as well as not completely independent.

The main conceptual difference of COP is that instead of introducing a separate programming construct, it generalizes classes. The injection direction in COP is aligned with the inclusion hierarchy (which generalizes inheritance) (Fig. 5). AOP allows for injecting behavior in any class of the program while COP restricts the injection area by child concepts only. In this sense, COP is more restrictive than AOP but it is also conceptually much simpler. We treat this as an advantage because the freedom provided by AOP has some significant drawbacks in the case of complex systems with a large number of aspects and classes. The possibility to modularize cross-cutting concerns in COP is a natural property of the existing mechanism of dual methods rather than a new independent mechanism. In AOP, the goal is reached by optional use of aspects while in COP it is reached by optional use of concepts (generalized classes) instead of classical classes.

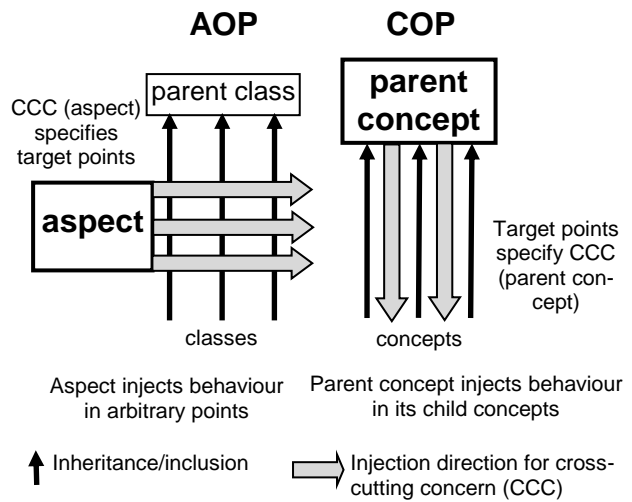

**Fig. 5.** Direction of injection in AOP and COP

Another important conceptual difference between COP and AOP is that they use the opposite directions in declaring join points (where some behavior has to be injected). In AOP, the behavior to be injected and the join points are defined within one aspect. As a consequence, the target join points are unaware of what kind of behavior will be injected into them. In COP, it is the target join point (child concept) that specifies what kind of cross-cutting behavior it needs by declaring its parent concept. In other words, if we include a concept in another concept then we declare that our methods will be wrapped into the corresponding functions provided by the parent concept. Such logic is a consequence of a more general principle that elements of a container or set inherit its properties while this contain can actively modify the behavior of all its members. The latter is treated as injection in terms of AOP.

**6.3 Incoming methods vs. inner methods**

"Beta is probably the most underappreciated object-oriented language in the world" [Ung09] and one of its major achievements is that it implemented the idea of treating subclasses as behavioral extensions to their superclasses. In other words, superclasses are supposed to provide generic behavior which is not overridden but really extended using the keyword inner [Kri87, Mad89]. Beta provides very clear and convincing justification for the need in such extension strategy (from least specific to most specific) which is also valid for COP. Imagine that the root class is a house consisting of many rooms which possibly have some interval division. To enter the house (parent object) we must go through its entrance door and then through the target room door to a target object. And only when we are inside, it is possible to do more specific actions. It would be unnatural (and in many cases completely unacceptable) to allow for direct access to the house internals bypassing the entrance and room doors. Yet, it is precisely how object-oriented inheritance work (with some exception like Beta) which can be characterized as the world without borders.

Beta was the first language aimed to fix this conceptual flaw. COP pursues the same goal but does it differently. COP introduces *two* kinds of methods, incoming and outgoing, which are accessible only from outside and inside, respectively. Incoming methods define a downward propagation direction using the sub keyword while outgoing methods define an upward direction using the super keyword. Thus two keywords and two propagation direction are made completely symmetric.

One of the reasons why inner methods have not been widely recognized is that they do not use object hierarchies and hence the protecting role of parent objects is not very actual. (Inner methods would be much more useful in prototype-based programming.) Since COP assumes that elements exist in a hierarchy, the mechanism of incoming methods becomes very natural and even necessary.

**6.4 COP vs. prototype-based programming**

The conception of re-use makes sense only if data or functionality can be *shared* among many elements. The most wide spread approach to sharing (and hence reuse) is based on using classes where child classes share their parent class. Yet, at the level of instances, only behavior can be shared while data is not shared and any child element gets its own copy of the parent. Thus the conception of reuse is not completely implemented in class-based approaches because objects still exist in flat space (Fig. 6a). Prototype-based programming (PBP) [Bor86; Lal86; Lie86] fixes this problem by allowing objects to exist in a hierarchy which results in many advantages including a new treatment of inheritance. However, to achieve this goal, PBP sacrifices classes by losing their obvious benefits (Fig. 6b). The main idea of PBP is that the behavior defined in an object is shared among all its child objects and hence an object is viewed as a standard example instance, called a prototype or exemplar. An object can inherit from and delegates to its prototype [Ste87].

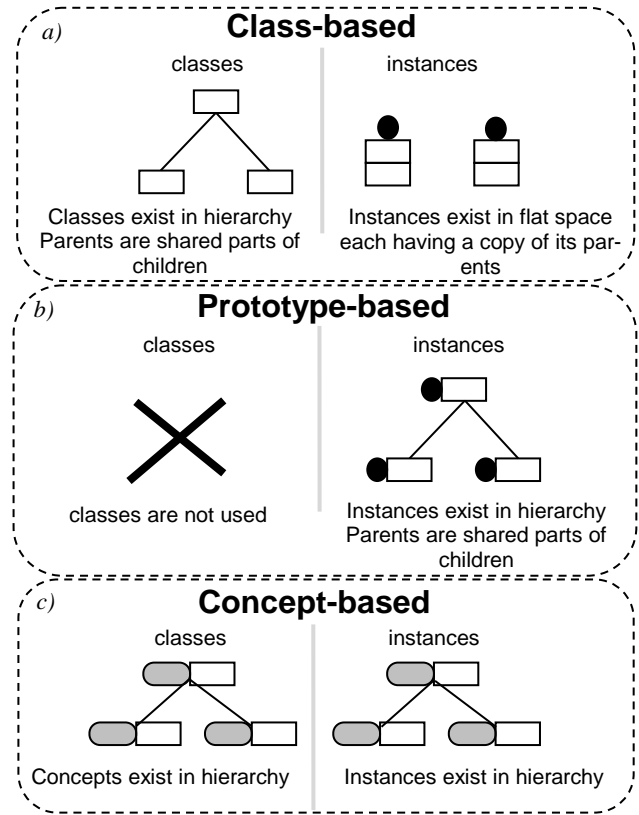

**Fig. 6.** COP unites class-based and prototype-based approaches

COP also assumes that objects exist in a hierarchy and "parents are shared parts of objects" [Cha91]. The difference between them is how this hierarchy is built and how it is interpreted. Object hierarchy in COP is built upon the notion of relative domain-specific reference which is implemented via value extension. In contrast, object hierarchy in PBP is viewed as a tree of objects represented by primitive references. Object references in PBP are still conventional reference as they are used in other object-oriented approaches. This means that objects exist in a hierarchy while their references exist in flat space. In COP, a reference is a complex value consisting of several segments each having an arbitrary structure. Thus both references and objects exist in a hierarchy (Fig. 6c). This hierarchy of reference-object couples is modeled by concepts (generalized classes). In this sense, COP combines class-based and prototype-based approaches. Another difference is that PBP uses delegation only in one direction: from children to parents. In contrast, COP introduces the opposite direction for delegation where parents can delegate to their children as described in Sections 5.2 and 6.3.

### 6.5 Conceptual modeling and data modeling

COP follows the Scandinavian tradition [MadMol88, Knud88] which views programming as a process of modeling [Ost90, Kri94]. However, COP not only supports main abstractions and relationships used in the object-oriented paradigm but also generalizes them by adding new treatments to the conventional notions like class and inheritance.

One novel layer of conceptual design which is supported in COP is *identity modeling* which is aimed at describing address spaces where (future) objects are supposed to exist. In COP, objects can be located anywhere (not even necessarily in a computer system) just because they exist in a *virtual* address space modeled by concepts. Conventional OOP also hides object references so that objects exist in a virtual space. However only primitive references are used for *all* objects. Hence there is no possibility to define new domain-specific address spaces with arbitrary access methods. Most existing object-oriented programming languages start from defining object classes without directly asking the question about the space where they will exist – it is assumed that it has to be done by the compiler. COP changes this rule and assumes that the first question in conceptual design is defining the space (home) for objects by describing their address structure and, even more important, the behavior of these addresses (references). And only after that objects can be defined.

Another novel conceptual feature of COP is that it directly supports containment relationship – one of the main abstractions in conceptual modeling and in mathematics in general. As a result, an object in COP is not an isolated instance like in most other approaches – it is inherently a set. Thus programming (that is, modeling) is reduced to describing sets and their members by specifying how objects are composed of other objects. Importantly, containment is not supported by mechanically adding a new construct or relationship but rather by generalizing inheritance. Thus conceptually, inheritance ('IS-A') in COP is a particular case of inclusion ('IS-IN') relation which is a new and rather strong assumption having very interesting consequences. To include an object into a container (a parent object) means to inherit its properties.

COP is an integral part of a novel general-purpose data model, call concept-oriented model (COM) [Sav09b, Sav11b, Sav12b, Sav14b], and the corresponding concept-oriented query language [Sav11a, Sav14a]. Shortly, COM can be viewed as COP plus partial order relation among objects. In particular, COM uses concepts and inclusion as they are defined in COP but in addition it is assumed that all objects are partially ordered using the principle that elements reference their greater elements. This partial order relation leads to significant simplification of the model by unifying many existing data modeling patterns. It can be interpreted in terms of multidimensional spaces [Sav05b], containment relation (inclusion by-reference) or as relationships among elements (as they are understood in the entity-relationship model). Partial order is also an important part underlying such mechanisms as logical navigation [Sav05c] and inference [Sav06].

## 7. Conclusion

In this paper we described a new version of concept-oriented programming. It revisits some classical notions like class, inheritance and reference, and therefore can be viewed as a generalization and further development of object-oriented programming. In particular, we argue that modeling values is as important as modeling objects and therefore both types of elements have to be supported in programming languages. We propose to use concepts for directly modeling values which are also references. Class as it is used in OOP is a particular case of concept. The focus in programming is now shifted in the direction of modeling how objects are represented and how they are accessed. Thus COP can be viewed as traditional OOP plus support of values and references.

In the case of concepts, extension operation gets a new treatment: it now describes an inclusion hierarchy while inheritance is viewed as a particular case of inclusion. Elements in a concept-oriented program exist in a hierarchy where each of them has a unique domain-specific reference. Such a hierarchy in traditional OOP can be traversed in only one direction using super-method calls. COP adds the opposite direction for navigating through the hierarchy using sub-method calls. Thus COP can be viewed as consisting of two parts: traditional OOP bottom up view (from children to parents) and the new top down (from parents to children) view. We have the following major assumptions which distinguish COP from OOP:

- Objects have to exist in a hierarchy which is modeled by their concepts (a flat space of objects is a particular case)

- Objects must have explicit references because references are as important as objects and also have structure and behavior. Having structure and behavior in only objects (represented by primitive references) is a particular case.

- Sub keyword is needed in order to delegate method calls down along the hierarchy. Using only super-method calls is a particular case.

- Parents have to be able to intercept requests to their children so that parent methods should have precedence over child methods. The classical view is that child methods have precedence over parent methods.

COP allows us to rethink several classical programming patterns by unifying and implementing them using only these principles:

- IS-A is a particular case of IS-IN. These two relations have always existed separately. In COP, to exist within some element means to inherit its properties and to be more specific element.

- Reference as a proxy. It is a wide spread pattern in C++ used to implement smart pointers and other intermediate functions. In COP, it can be viewed as one of the main patterns: describe address space and access methods before the elements in this space. In other words, what happens during access is more important than what happens in end points.

- Aspect-orientation and cross-cutting concerns. COP allows for explaining this phenomenon in terms of object hierarchies without using additional independent constructs. Thus programming in COP is always "aspect-oriented" just because parent concepts define functionality which is guaranteed to

- Polymorphism and overriding. Classically, more specific elements are supposed to have precedence over more general elements. This principle exists in various forms and probably nobody doubts that it can be different. Yet, we also rethink it and postulate that parents have priority over children when processing incoming calls (for external requests).

In summary, COP can be viewed as a generalization of OOP because it retains its main features by adding possibilities to model references, access procedures and object hierarchies. It also allows for modularizing cross-cutting concerns in a novel OOP-compatible way by unifying aspect-based and class-based approaches. COP is also compatible with prototype-based programming because it also thinks of objects as existing in a hierarchy and uses delegation mechanisms for implementing inheritance. COP also is very close to conceptual modeling and data modeling. Taking these properties into account, COP in its current form can be used as a basis for a next generation programming model.